\documentclass[10pt,journal,compsoc]{IEEEtran_jnal}

\usepackage{graphics}
\usepackage{graphicx}
\usepackage{amsmath}
\usepackage{amsfonts}
\usepackage{amssymb}%
\usepackage[linesnumbered,boxed]{algorithm2e}
\usepackage{cases}
\usepackage{xspace}
\usepackage{multirow}
\usepackage{slashbox}
\usepackage{subfigure}
\usepackage{listings}
\usepackage{color}
\usepackage{url}
\usepackage{ftnxtra}
\usepackage{pbox}
\usepackage{color}
\usepackage{ragged2e}

\newcommand{\eg}{\textit{e.g.},\xspace}
\newcommand{\etal}{\emph{et al.}\xspace}

\makeatletter
\renewcommand{\paragraph}{\@startsection{paragraph}{4}{\z@}
	{4pt plus 6pt}{\z@}{\normalsize\itshape}}
\makeatother

\ifCLASSINFOpdf
\else
\fi

\begin{document}
\title{User Review-Based Change File\\ Localization for Mobile Applications}

\author{Yu Zhou, Yanqi Su, Taolue Chen, Zhiqiu Huang, Harald Gall, Sebastiano Panichella
\IEEEcompsocitemizethanks{\IEEEcompsocthanksitem Yu Zhou, Yanqi Su and Zhiqiu Huang  are with College of Computer Science and Technology,
Nanjing University of Aeronautics and Astronautics, Nanjing, China. Yu Zhou is also with the State Key Laboratory for Novel Software Technology, Nanjing University, Nanjing, China. \protect\\
E-mail:  \{zhouyu,suyanqi,zqhuang\}@nuaa.edu.cn
\IEEEcompsocthanksitem T. Chen is with Department of Computer Science and Information Systems, Birkbeck, University of London, UK. He is also with the State Key Laboratory for Novel Software Technology, Nanjing University, Nanjing, China. \protect\\
E-mail: taolue@dcs.bbk.ac.uk
\IEEEcompsocthanksitem H. Gall is with Department of Informatics, University of Zurich, Switzerland.\protect\\
E-mail:  gall@ifi.uzh.ch
\IEEEcompsocthanksitem S. Panichella  is with Zurich University of Applied Science, Switzerland.\protect\\
E-mail:  panc@zhaw.ch
}
}

\IEEEtitleabstractindextext{
{\justifying
\begin{abstract}
In the current mobile app development, novel and emerging DevOps practices (e.g., Continuous Delivery, Integration, and user feedback analysis) and tools are becoming more widespread. For instance, the integration of user feedback (provided in the form of user reviews) in the software release cycle represents a valuable asset for the maintenance and evolution of mobile apps. To fully make use of these assets, it is highly desirable for developers to establish semantic links between the user reviews and the software artefacts to be changed (e.g., source code and documentation), and thus to localize the potential files to change for addressing the user feedback. In this paper, we propose \textsc{RISING} (\textbf{R}eview \textbf{I}ntegration via cla\textbf{S}sification, cluster\textbf{I}ng, and linki\textbf{NG}), an automated approach to support the continuous integration of user feedback via classification, clustering, and linking of user reviews. \textsc{RISING} leverages domain-specific constraint information and semi-supervised learning to group user reviews into multiple fine-grained clusters concerning similar users' requests. Then, by combining the textual information from both commit messages and source code, it automatically localizes potential change files to accommodate the users' requests. Our empirical studies demonstrate that the proposed approach outperforms the state-of-the-art baseline work in terms of clustering and localization accuracy, and thus produces more reliable results.
\end{abstract}
}
}

\maketitle
\IEEEdisplaynontitleabstractindextext
\IEEEpeerreviewmaketitle

\IEEEraisesectionheading
{\section{Introduction}\label{intro}}

\IEEEPARstart{T}{he} extensive proliferation of smart devices represents one of the most visible technology and society advances of the last years. Indeed, mobile phones, tablets and smart watches are widely used in many aspects of today's life \cite{Dehghani16,Martin:tse2017}. This phenomenon is particularly reflected in the growth of the app industry, with millions of developed and maintained mobile applications \cite{statista,business}.

This trend also impacts  the current mobile app development, which is characterized by novel and emerging DevOps practices (e.g., Continuous Integration, Deployment, Deliver, and user feedback analysis) and tools \cite{devops-mobile-apps1,devops-mobile-apps2}. For instance, the integration of user feedback (provided in the form of user reviews) in the software release cycle represents a valuable asset for the maintenance and evolution of these apps \cite{Harman:MSR:2012,panichella2015can}, or for ensuring a reliable testing automation for them \cite{Grano:20172}. Thus, a key and winning aspect of successful apps is related to the capability of developers to deliver high-quality apps and, at the same time, address user requests; this is crucial for the app to stay on the market and to keep gaining users \cite{Martin:tse2017, Machiry:2013}.  

Mobile user reviews, mainly appear in major online app stores (e.g., Google Play and Apple AppStore), provide valuable feedback for further improvements of mobile apps. They might report software bugs, complain about usage inconvenience, request new features, etc \cite{Dehghani16,panichella2015can,Chen:2014:AMI:2568225.2568263}. Such information is valuable for developers, since it represents crowd-sourced knowledge from the customers' perspective,
providing useful information for the evolution and release planning of mobile apps~\cite{Harman:MSR:2012,panichella2015can,Chen:2014:AMI:2568225.2568263}. As a concrete example, among many reviews of a popular instant messenger app signal\footnote{\scriptsize https://play.google.com/store/apps/details?id=org.thoughtcrime.securesms}, one group concentrates on the theme issues. Particularly, one review states that ``\emph{Wish it had a dark or black theme}." In the following release of the app, new themes, including the aforementioned dark and black ones, were integrated. As another example, for the app AcDisplay\footnote{\scriptsize https://play.google.com/store/apps/details?id=com.achep.acdisplay}, one review states ``\emph{It would be great if you could just use the proximity sensors to wake the screen much like the Moto app uses IR sensors when you wave over the phone}." Later on this feature was added in the next version of the app.

Due to the high number of app user reviews developers receive on a daily basis
(
popular apps could receive more than 500 reviews per day on average~\cite{mcilroy17}),  collecting and analyzing them manually becomes increasingly infeasible.  As a result, developers are interested in adopting automated approaches which are 
able to classify/cluster such reviews, and to localize potential change files.  This is key to enhance the development productivity, and in turn, to facilitate the continuous delivery of app products.

Recent work has proposed tools for user feedback classification \cite{panichella2015can, ciurumelea2017analyzing}, clustering \cite{villarroel2016release, palomba17} and summarization \cite{sorbo16fse, SorboPAVC17}. 
Unfortunately, most of these tools suffer from various important limitations.
First, the classification or clustering accuracy  is 
hindered by the general low-quality of user reviews
\cite{Chen:2014:AMI:2568225.2568263,panichella2015can,palomba17}. Compared to other kinds of software artefacts such as software documents, bug reports, logging messages which are provided by developers, reviews are generated by (non-technical) users, who tend to produce reviews of lower-quality (e.g., the textual descriptions are usually short and unstructured, mixed with typos, acronyms and even emojis~\cite{vu2015}).
Second, existing  classification and clustering is usually conducted at a coarse-grained sentence level containing potentially multiple topics without taking domain-specific knowledge into account. This further reduces the accuracy of classification/clustering, and impedes the effectiveness of further localization.
Third, and the most important, available tools are not able to cope with the lexicon gap between user reviews and software artefacts (e.g., the source code) of the apps, which makes the standard textual similarity based localization approaches less effective \cite{palomba17,Grano:20172,ciurumelea2017analyzing}.
Consequently 
existing tools 
have to settle with a low file localization accuracy \cite{palomba17,ciurumelea2017analyzing}. 

To overcome the aforementioned limitations, in this paper, we propose \textbf{\textit{RISING}} (user-\textbf{R}eviews \textbf{I}ntegration via cla\textbf{S}sification, cluster\textbf{I}ng, and linki\textbf{NG}),
an automated approach to support the continuous integration of user feedback via classification, clustering, and linking of user reviews.
Specifically, \textit{RISING} leverages domain-specific constraint information and semi-supervised learning to group reviews into multiple fine-grained clusters concerning similar user requests. Then, by combining the textual information from both commit messages and source code, it automatically localizes the files to change to accommodate the users' requests. Our empirical studies demonstrate that the proposed approach outperforms state-of-the-art baselines \cite{palomba17} in terms of accuracy, providing more reliable results.

The main contributions of the paper are summarized as follows:
\begin{itemize}
\item We propose a semi-supervised clustering method by leveraging domain-specific knowledge to capture constraints between the reviews. The experiments demonstrate its efficacy in improving the accuracy of clustering, in particular, its superiority to other clustering methods reported in the literature. 
	
\item We propose a change file localization approach by exploiting commit messages as a media to fill the lexicon gap between user reviews and software artefacts, which, as the experiments demonstrate, enhances the localization accuracy substantially.
	
\item We collect user reviews and commit messages from 10 apps available from Google Play and Github, and prepare a dataset with the processed reviews and commit logs.\footnote{https://csyuzhou.github.io/files/dataset.zip} This will not only facilitate the replication of our work, but also serve other related software engineering research for example mining mobile app store and intelligent software development.
\end{itemize}



\medskip
\noindent \textbf{Structure of the paper.} Section~\ref{bg} gives background information for a better understanding of the context of our work, while Section~\ref{approach} details the proposed approach to address the limitations of state-of-the-art approaches on user reviews analysis. Section~\ref{case}  presents the main research questions driving our investigation and describes the case studies we conduct to answer the research questions. In Section~\ref{discuss} we provide some discussions, including the threats that may bias our results and how we mitigate them. Related work is discussed in Section~\ref{related}, while Section~\ref{conclusion} concludes the paper and describes our future research agenda.

\section{Background}\label{bg}
This section provides a brief overview  of (i) the contemporary development pipeline of mobile applications and 
(ii) the importance of user feedback analysis in the mobile context. Section~\ref{related} complements this section by providing related work on user feedback analysis and applying Information Retrieval (IR) in software engineering, with a specific emphasis on the mobile application domain.

\begin{figure*}[t]
\centering
	\includegraphics[width=0.85\textwidth]{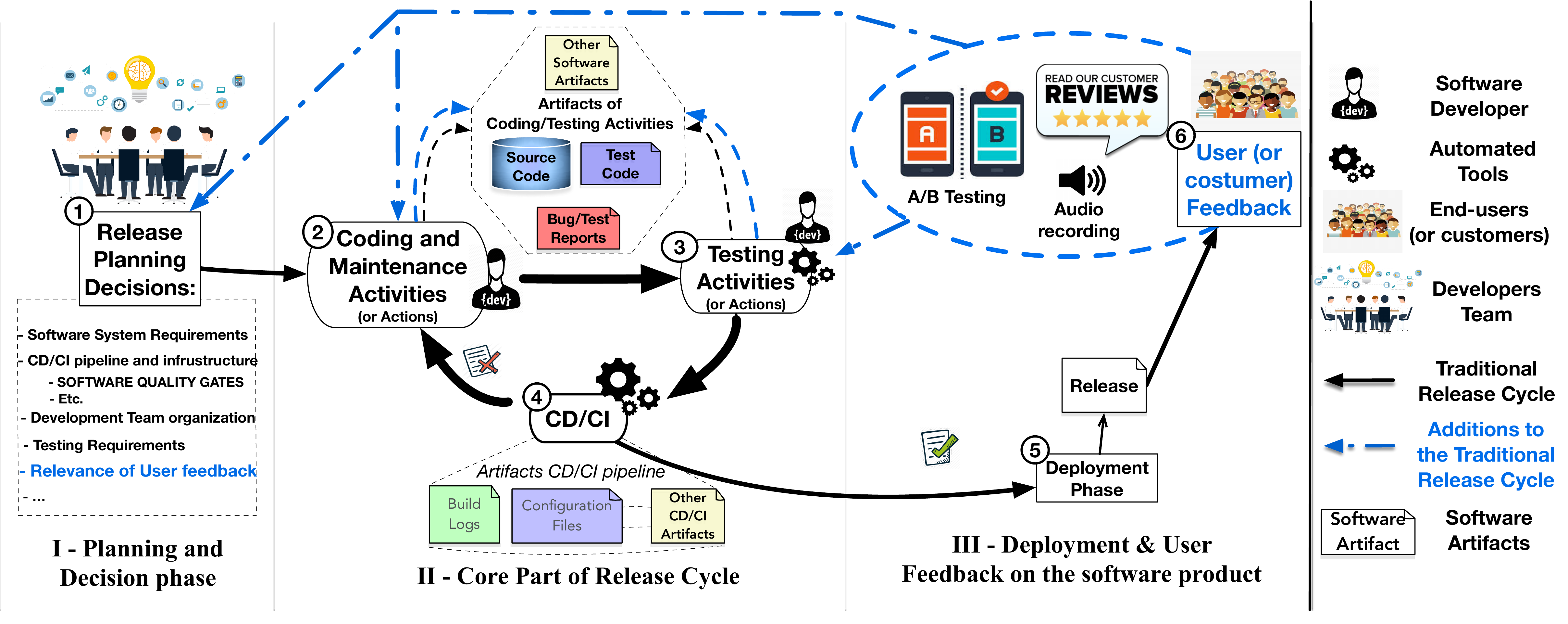}
		\vspace{-3 mm}
	\caption{
	Release Cycle}
	\label{fig:overview}
	\vspace{-3 mm}
\end{figure*}

\medskip
\noindent\textbf{Development Release Cycle of Mobile Apps}. As shown in Figure~\ref{fig:overview} \cite{beck2001agile}, the conventional mobile software release cycle has evolved in recent years into a more complex process, integrating DevOps software engineering practices \cite{Duvall:2007,LaukkarinenKM17}. The DevOps movement aims at unifying the conflicting objectives of software development (Dev) and software operations (Ops),  with tools for  shortening release cycle activities. Continuous Delivery (CD) is one of the most emerging DevOps software development practices, in which developers' source/test code changes are sent to server machines  to automate all software integration (e.g., building and integration testing) tasks required for the delivery \cite{Humble:2010}. When this automated process fails (known as \textit{build failure}), developers are forced to go back to coding to discover and fix the root cause of the failure \cite{IslamZ17a,ZiftciR17,VassalloSZRLZPP17}; otherwise, the changes are released to production in short cycles. These software changes are then notified to users as new updates of their mobile apps. In this context, users usually provide feedback on the new version (or the A/B testing versions) of the apps installed on their devices, often in the form of comments in app reviews \cite{Chen:2014:AMI:2568225.2568263,panichella2015can,DiSorbo:2016:UCM:2950290.2950299,villarroel2016release,SorboPAVC17}. For completeness, it is important to mention that developers do not look only into user reviews to gather information about the overall user satisfaction. Indeed, they also rely on metrics concerning users' behavior inside the app \cite{8606261,NayebiFR17}, and this is especially true  when they apply A/B testing strategies \cite{NayebiAR16}.

\medskip
\noindent\textbf{User Feedback Analysis in the Mobile Context}.
Mobile user feedback stored in different forms (e.g., user reviews, videos recorded, A/B testing strategies,  etc.), can be used by developers to decide possible future directions of development or maintenance  activities \cite{panichella2015can,noei18}. Therefore, user feedback represents a valuable resource to evolve software applications  \cite{DiSorbo:2016:UCM:2950290.2950299}. 
As a consequence, the mobile development would strongly benefit from integrating User Feedback in the Loop (UFL) of the release cycle \cite{Grano:20172,NagappanS16,Grano:2017,pelloni2018becloma,Panichella18} (as highlighted by the {\textbf{\textit{blue}}} elements/lines shown in Fig.~\ref{fig:overview} \cite{beck2001agile}), especially on the testing and maintenance activities. This has pushed the software engineering research community to study more effective automated solutions to \textit{``enable the collection and integration of user feedback information in the
development process''} \cite{NagappanS16}. The key idea of the techniques for user feedback analysis is to model \cite{Chen:2014:AMI:2568225.2568263,panichella2015can,DiSorbo:2016:UCM:2950290.2950299,ciurumelea2017analyzing},  classify \cite{Chen:2014:AMI:2568225.2568263,panichella2015can,ciurumelea2017analyzing}, summarize \cite{Panichella18,DiSorbo:2016:UCM:2950290.2950299,SorboPAVC17} or cluster \cite{villarroel2016release} user feedback in order to  integrate them into the release cycle. The research challenge is to effectively extract the useful feedback  to actively  support the developers to accomplish the release cycle tasks.

\medskip
\noindent\textbf{Mobile Testing and Source Code Localization based on User Feedback Analysis}.
User feedback analysis can potentially provide to developers information about the changes to perform to achieve a better user satisfaction and mobile app success. However, user reviews analysis alone is not sufficient to concretely help developers to continuously integrate user feedback information in the release cycle, and in particular (i) maintenance \cite{ciurumelea2017analyzing,palomba17,panichella2015can,DiSorbo:2016:UCM:2950290.2950299} and (ii) testing \cite{Grano:2017,Grano:20172,Panichella18,pelloni2018becloma} activities.
Recent research directions push the boundaries of user feedback analysis in the direction of  \textit{change-request file localization} \cite{ciurumelea2017analyzing,palomba17} and \textit{user-oriented testing}  (where user feedback is systematically integrated into the testing process) \cite{Grano:20172,Grano:2017,pelloni2018becloma}. We will elaborate more \textcolor{red}{on} the literature in Section~\ref{related}.

In this paper we focus on supporting developers with more advanced and reliable approaches to derive and cluster change-requests from user feedback, thus localizing the  files to change \cite{ciurumelea2017analyzing,palomba17}  to better support mobile maintenance tasks \cite{ciurumelea2017analyzing,palomba17,panichella2015can,DiSorbo:2016:UCM:2950290.2950299}.

\section{Approach}\label{approach}

\begin{figure}[h]
	\centering
	\centering
	\includegraphics[width=9.0cm, height=5.6cm]{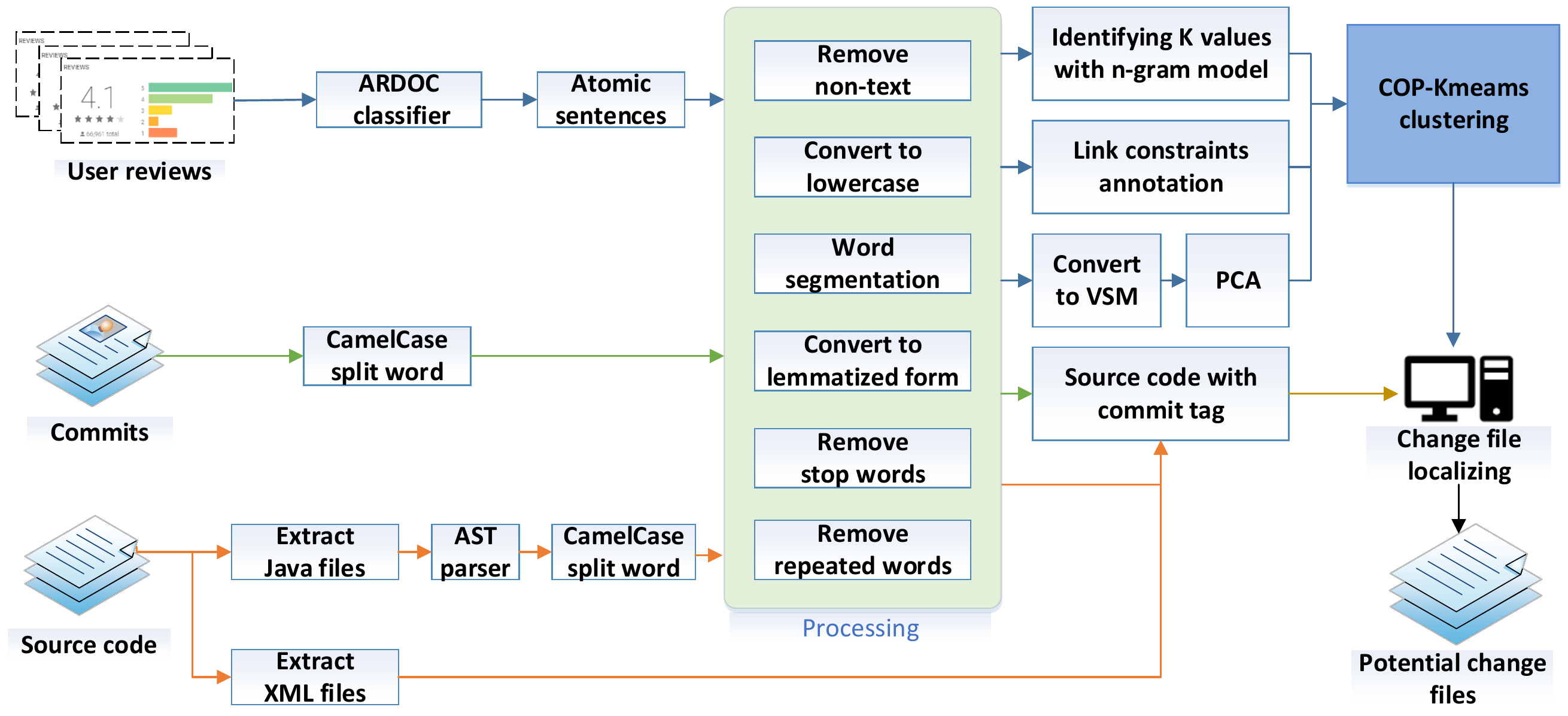}
	\vspace{-3mm}
	\caption{Approach Overview} \label{fig:framework}
\end{figure}

As have been identified in the introduction, there are three major limitations in the existing approaches, i.e., low accuracy of classification and clustering because of low-quality of user reviews, difficulties in coping with the different vocabularies used to describe users' experience with the apps, and the existence of the lexicon gap between user reviews and software artefacts. RISING employs various techniques to mitigate these issues on which we will elaborate in this section. Outlined in  Fig.~\ref{fig:framework}, RISING consists of two major parts, i.e., \emph{clustering} and \emph{localization}. In the first branch,  user reviews go through a series of textual processing, including fine-grained review segmentation 
, non-text removal, and lemmatization conversion; in the second branch, source code is first tagged with the commit messages, and comprehensive similarity scores are to be calculated, based on which the localization recommendations will be made. The details of these two parts will be given in Section~\ref{sec:cluster} and Section~\ref{sec:localization} respectively.

\subsection{User review clustering} \label{sec:cluster}
Most user reviews are short textual snippets  consisting of multiple sentences. These raw sentences may address different aspects of apps and need to be preprocessed before clustering. Based on their content, the reviews can mainly be  classified into four categories, i.e., information giving, information seeking, feature request and problem discovery~\cite{panichella2015can,palomba17}. In particular, ``Information giving" denotes those sentences that inform or update users or developers about an aspect related to the app; ``information seeking" denotes those which attempt to obtain information or help; ``feature request" denotes those expressing ideas, suggestions or needs for improving the app's functionalities and performance; ``problem discovery" denotes the sentences describing issues with the apps or their unexpected behaviors~\cite{panichella2015can}.
Since our aim is to identify those reviews which are directly relevant to apps' evolution, following \cite{palomba17} we only focus on the last two categories, i.e., feature request and problem discovery. To this end, we first employ ARDOC, a \emph{user review classifier} developed in the previous work~\cite{panichella2015can} which  transforms user reviews into individual sentences and then classifies these sentences into one of the aforementioned four categories. We then collect those sentences of the last two categories. ARDOC is built upon the functions of AR-Miner~\cite{Chen:2014:AMI:2568225.2568263}, which can filter noisy and uninformative user reviews in the first place. Such capability contributes another benefit to our approach.

To improve the accuracy of clustering, two tactics are employed, i.e., finer granularity review segmentation and textual processing, which will be elaborated in the following two subsections.

\subsubsection{Fine-grained review segmentation}
Clustering user reviews 
is usually conducted at the sentence level. 
We observe that, even inside an individual sentence, there still may  be multiple topics involved which possibly address quite different concerns. As an example, one user review of AcDisplay reads ``I wish there was a pattern lock feature and a camera shortcut for the lockscreen.''  Apparently, the user prefers two more features (a pattern lock and a shortcut utility). Moreover, for composite sentences in user reviews, if they contain adversative conjunctions such as 'but', the content after `but' usually discloses the real information. As an  example from K9-Mail\footnote{https://play.google.com/store/apps/details?id=com.fsck.k9}, one user states that ``This app is good, but it is lacking a key feature for anyone who uses mailing lists: Reply-To-List.'' In this case, for the purpose of localization, the content before `but' is not  informative at all, and may introduce noises to the follow-up process. As a result, we propose to have a more fine-grained text analysis. In particular, we split the composite sentences into \textsc{atomic} ones each of \textit{which expresses a single concern only}, and remove the irrelevant part of the sentence.


To achieve that, we employ a statistical parser from the Stanford NLP toolkit\footnote{https://nlp.stanford.edu/software/lex-parser.shtml} to generate grammatical structures of sentences, i.e., phrase structure trees. We then traverse the leaf nodes of the phrase structure tree to determine whether or not the sentence contains conjunctions. Particularly, we focus on two types of conjunctions, i.e., copulative conjunctions and adversative conjunctions. The former (e.g., `and', `as well as' and `moreover') mainly expresses the addition while the latter (e.g.,  `but', `yet') denotes contrasts.

For the first type, we recursively parse the nodes to identify the layer where the copulative conjunctions are located. We then obtain the copulative conjunction's sibling nodes. The two parts connected by the conjunction may be two sentences, two noun phrases, two verb phrases, etc. Given different conditions, we can generate two atomic sentences based on the parts which are connected by the conjunctions. As a concrete example, if the conjunction `and' connects two noun objectives, then the two objectives are split as the only objective of each atomic sentence, but they share the same subjective and verb. (e.g. I wish there was a pattern lock feature and a camera shortcut for the lockscreen. $\rightarrow$ I wish there was a pattern lock feature for the lockscreen. I wish there was a camera shortcut for the lockscreen). If the conjunction 'and' connects two sentences, then the two sentences will be simply split into two atomic sentences (e.g. There are only 2 things I'd change for a 5 star review; I wish it had audio controls, and I wish there was a camera shortcut from the lock screen. $\rightarrow$ There are only 2 things I'd change for a 5 star review; I wish it had audio controls. There are only 2 things I'd change for a 5 star review; I wish there was a camera shortcut from the lock screen).

For the second type, since we believe that the content after the adversative conjunctions convey the real information, we only preserve the leaf nodes after the conjunction nodes and simply leave out the other parts.

\subsubsection{Textual processing}
User reviews are generally informal and unstructured, mixing with typos, acronyms and even emojis~\cite{vu2015}.
The noisy data inevitably degrades the performance of clustering and localization which necessitates further textual processing. 
We first filter out the emoji characters and other punctuation content. Some emojis which were published as icons are stored in a text format, and their encoding appears as combination of question marks. Some others also use a combination of common punctuations, such as smiley faces. These patterns are matched by using regular expressions. Particularly, we propose two regular expressions to extract the pattern. The first one is "$\backslash\backslash p\left\{P\right\}\backslash\backslash s ^\ast$". It removes all punctuations and replaces them with a space; the second one is "$\left[^\wedge a-zA-Z0-9\backslash\backslash s\right] ^\ast$" which removes non-alphanumeric parts. Furthermore, we also convert all letters to lowercase uniformly.

Given the above steps, sentences are transformed into lists of words (i.e., tokens). We then use the Stanford NLP toolkit\footnote{https://stanfordnlp.github.io/CoreNLP/} to transform the inflected words to their lemmatization form. Here a dictionary-based  instead of a rule-based approach is used to convert words into tokens which can avoid  over-processing of words. (For instance, ``images'' is transformed correctly to image instead of to imag).
User reviews may contain stopwords that could introduce noise for clustering and need to be removed. We note that the existing English stopword list cannot be well applied here for two reasons: first, a large number of user reviews contain irregular acronyms (e.g., asap--as soon as possible, cuz--cause) which cannot be processed by the existing stopword list. Second, some words are in the regular stopword list, but for specific apps, they may convey important information. For example, some words, such as ``home'', listed in strings.xml which encodes the string literals used by the GUI components, are of this kind. Therefore, we manually edit the English stopword list\footnote{The customized stopword list is also available online with the replication package.} accordingly (e.g., by adding some acronyms commonly used and removing some words that appear in strings.xml). We also delete the repeated words and the sentences which contain less than two words, because in short documents like user reviews, documents with less than two words hardly convey any useful information for evolution purpose.

Note that review segmentation is executed before textual processing, because the textual processing, which includes transforming the inflected words to their lemmatization form, removing stopwords, etc, would affect the grammatical structures of sentences which are crucial for review segmentation.


\subsubsection{User Review Clustering}

Although ARDOC could classify reviews into  ``problem discovery'' and ``feature request'', such coarse-grained classification provides limited guidance for developers when confronted with specific maintenance tasks. A more fine-grained approach is highly desirable. Firstly, it is not uncommon that the number of user reviews makes addressing every concern  practically infeasible. Therefore, developers would like to identify the most common issues or requests raised by the end users, which are supposed to be treated with higher priority~\cite{villarroel2016release}. Secondly,  not all user reviews are meaningful, especially in the problem discovery category. In practice, some complaints are actually caused by users' misunderstanding. By grouping similar issues together, such cases would be easier to be identified. Both of these motivate using clustering of pre-processed user reviews.

\medskip
\noindent\textbf{Construction of word-review matrix.}
We adopt the widely used Vector Space Model (VSM~\cite{baeza1999modern}) to represent the pre-processed texts.
We fix a vocabulary $\Sigma$, each of which represents a feature in our approach.
Let $n=|\Sigma|$, the size of the vocabulary, and $m$ be the number of atomic sentences. We first construct a raw matrix $WR_{m\times n}$ where each entry $WR[r,w]$ is equal to the number of occurrences of the word $w$ in the review $r$.

For each word $w\in \Sigma$, let $f_w$ denote the occurrence of $w$ in all reviews, i.e., $f_w:=\sum_{r} WR[r, w]$,  
and we use logarithmically scaled document frequency ($df(w)$) as the weight assigned to the corresponding word:
\[
df(w) = \log (1 + f_{w})
\]

Finally we can construct the scaled word-review matrix $R_{m\times n}$, where each entry
\[ R[r,w]:= WR[r,w]*df(w).\] 

We remark that some related work uses traditional tf/idf as the weighting strategy~\cite{villarroel2016release,adelina17}. However, we use the document frequency (df)  \cite{baeza1999modern} for two reasons: (1) Clustering in our approach is done at the sentence level. Particularly, these sentences are short where an individual word usually occurs once, so tf would be  meaningless for clustering in most cases.
(2) In general, the purpose of idf is to give less weight to common words than to less common ones. In our preliminary experiments, we found that some words which only appear once do not make right suggestion to developers because they only represent personal feedback/opinion and lack a general ground. However, they carry high idf simply because they are rare. On the other hand, those words which can indeed represent common issues encountered, or new functions required, by a majority of users carry low weights. To offset, we adopt df rather than more common tf/idf. Besides, one of the strong reasons to use idf is to reduce the weight of stop words which would have been removed in the data preprocessing steps.

Due to the large number of user reviews and the shortness nature of individual atomic sentences, the word vectors are of very high-dimension but very sparse. To reduce the dimension, we use the principal component analysis (PCA) technique~\cite{cohen14,ding04} which is one of the most widely used techniques for dimension reduction. Essentially, PCA replaces the original $n$ features with an (usually much smaller) number $r$ of features. The new features are linear combinations of the original ones that maximize the sample variance and try to make the new $r$ features uncorrelated. The conversion between the two feature spaces captures the inherent variability of the data. Finally, the resulting matrix of $m\times r$ dimension gives rise to the data set $D$, which is the collection---as vectors---of rows of the matrix.

\medskip
\noindent\textbf{COP-Kmeans.}
After obtaining the vector models, we are in position to cluster similar texts based on their content. Existing approaches mainly employ automatic clustering algorithms to divide the reviews into multiple groups. However, we postulate that clustering would benefit from leveraging domain knowledge about the mobile app dataset. By investing limited human effort, the performance of clustering could be further boosted. For example, from AcDisplay, some reviews state ``I do wish you could set a custom background, though.'' and ``Would be nice to be able to customize the wallpaper too.'' As for traditional clustering algorithms, since the two keywords (i.e., background and wallpaper) are quite different in regular contexts, these two sentences would have a very low similarity score and thus be clustered into two different categories. However, professional developers would easily recognize that ``wallpaper'' and ``background'' refer to similar things in UI design, which suggests that the two reviews address the same issue and should be put into the same cluster.

On the other hand, some reviews might address quite irrelevant issues using the same words. For example, again in AcDisplay, two reviews are as below: ``I would love the option of having different home screen.'', and ``First I'd like to suggest to disable that home button action because it turns the lock screen off ..., I hope you do it in next update.''. These two reviews have completely different meanings, but since they both contain key words ``home'' and ``screen'', they are very likely to be clustered together by traditional clustering algorithms. 

Domain knowledge of developers could potentially improve the precision of clustering, which has not been 
exploited by the traditional 
clustering algorithms. To remedy this shortcoming, 
we annotate a subset of instances with two types of link information, i.e., must-link and cannot-link constraints, as a priori knowledge and then apply the constrained K-means clustering technique~\cite{wagstaff01}.
%
The  must-link constraints specify the instance pairs that discuss semantically similar or the same concerns, judged by professional developers with rich development expertise. Likewise, the cannot-link constraints specify the instance pairs that are \emph{not} supposed to be clustered together. 
Besides, the must-link constraints define a transitive binary relation over the instances~\cite{wagstaff01}.
When making use of the constraints (of both kinds), we take a transitive closure over the constraints. (Note that although only the must-link constraints are transitive, the closure is performed over both kinds because, e.g., if $d_i$ must link to $d_j$ which cannot link to $d_k$, then we also know that $d_i$ cannot link to $d_k$.)

To use the K-means family of algorithms, one needs to determine the value of the hyper-parameter $K$. There are some traditional, general-purpose approaches~\cite{hamerly03,tibshi00,pelleg00}, 
but they did not take the topic distribution concerns into consideration so cannot  
provide a satisfactory solution in our setting. We instead use a heuristic-based method to infer $K$. The heuristic is derived from the n-gram model of the review texts, since we believe the cluster number should strongly correlate  to the topic distribution. N-gram denotes a sequence of $n$ words in a particular sentence, which  is a widely adopted statistical model to predict the occurrence of the $n$-th word using its previous $n-1$ words,  based on the probability distribution of these words.
Concretely, we obtain the 2-gram phrases of all user reviews. Then we merge the same phrases and record the number of occurrences of those phrases. If two phrases share the same word information, the less frequent phrase will be deleted. We also delete the phrases which occur once. $K$ is then set to be the number of the remaining phrases. 
(2-gram is used as we empirically found that this yields the best performance.) 

The COP-Kmeans algorithm takes the must-link and cannot-link dataset, $K$ value and atomic sentence vectors as input and produces 
the clustering results. The pseudo-code 
is given in Algorithm~\ref{al1}. First, it randomly selects $k$ samples $\left\{\mu_1,\ldots,\mu_k\right\}$ from the data set $D$ as the initial cluster centers. Then, for each sample $x_i$ in $D$, assign it to the closest cluster $C_j$ such that 
it doesn't violate constraints in $M$ and $C$. If no such cluster exists, an error message (line 4-21) would be returned. Then, for each cluster $C_j$, update its centroid by averaging all of the points $x\in C_j$ (line 22-24). This process iterates until the mean vectors no longer change.

\begin{algorithm}
    \SetKwInOut{Input}{\textbf{Input}}\SetKwInOut{Output}{\textbf{Output}}

    \Input{
        \\
        The Data set $D = \left\{x_1,x_2,...,x_m\right\}$\;\\
        The Must-link constraints $M$\;\\
        The Cannot-link constraints $C$\;\\
        The K-value $k$\;\\}
    \Output{
        \\
        The clustering results$\left\{C_1,C_2,...,C_k\right\}$\;\\}
    \BlankLine

    Randomly select $k$ samples$\left\{\mu_1,\mu_2,...,\mu_k\right\}$ from $D$ as the initial cluster centers\;
    \Repeat
        {\text{Mean vectors are no longer updated}}
        {$C_j = \varnothing(1 \leq j \leq k)$\;
            \For {$i = 1,2,...,m$}{
                Calculate the distance between the sample $x_i$ and each mean vector $\mu_j(1 \leq j \leq k):d_{ij}= \left \|x_i-\mu_j \right \|_2$\;
                $\mathcal{K} = \left\{ 1,2,...,k\right\}$\;
                is\_merged = false\;
                \While{$\urcorner$ is\_merged}{
                    Find the cluster closest to the sample $x_i$ based on $\mathcal{K}$: $r = \arg\min_{j \in \mathcal{K}}d_{ij}$\;
                    Detecting whether $x_i$ is classified into cluster $C_r$ violates constraints in $M$ and $C$\;
                    \uIf{$\urcorner$ is\_voilated}{
                        $C_r = C_r \cup \left\{ x_i\right\}$\;
                        is\_merged=true\;
                    }
                    \Else{
                        $\mathcal{K} = \mathcal{K} \setminus \left\{ r\right\}$\;
                        \If{$\mathcal{K} = \varnothing$}{
                            \textbf{Break} Return error message\;
                        }
                    }
                }
            }
            \For {$j = 1,2,...,k$}{
                $\mu_j = \frac{1}{\left|C_j\right|}\sum\limits_{x\in C_j}{x}$\;
            }
        }
    \caption{Constrained K-means Algorithm\label{al1}}
\end{algorithm}


\subsection{Change file localization} \label{sec:localization}
%
%


For localizing potential change files, our approach combines the information from both the commit message and the source code. To get the commit messages of mobile apps, we exploit open-source projects  
to collect (i) the title, (ii) the description, (iii) the set of files involved, and (iv) the timestamp, for each commit. For source code, we mainly use the file path, class summary, method summary, method name and field declaration. Class summary and method summary can be extracted based on the javadoc tags. Method names and field declarations are parsed through abstract syntax tree (AST) analysis.
In both cases, we remove non-textural information, split identifiers based on camel case styles, convert letters to lower case formats, stem, and remove stopwords/repetition words.
Finally, the bag-of-words (BoW) model from the target app's source code and commit messages are generated respectively.





\subsubsection{Tag Source Code Files} \label{sect:tag}

As mentioned earlier, we propose to leverage historical commit information to bridge the semantics gap between user reviews and source code. To this end, we first tag the source code with the historical change information. Particularly, for each commit, we extract the title, description, timestamps, and the involved file paths. From the file paths, we traverse the corresponding source code files in the project, and all the collected information, i.e., the title, description, and time stamps, is attached with the source file. As a result, each source code file can be regarded as a pair,
\[file=(code, commit)\]
where both $code$ and $commit$ are bag of words.

\begin{figure}[h]
	\centering
	\centering
	\includegraphics[width=8.8cm, height=5.8cm]{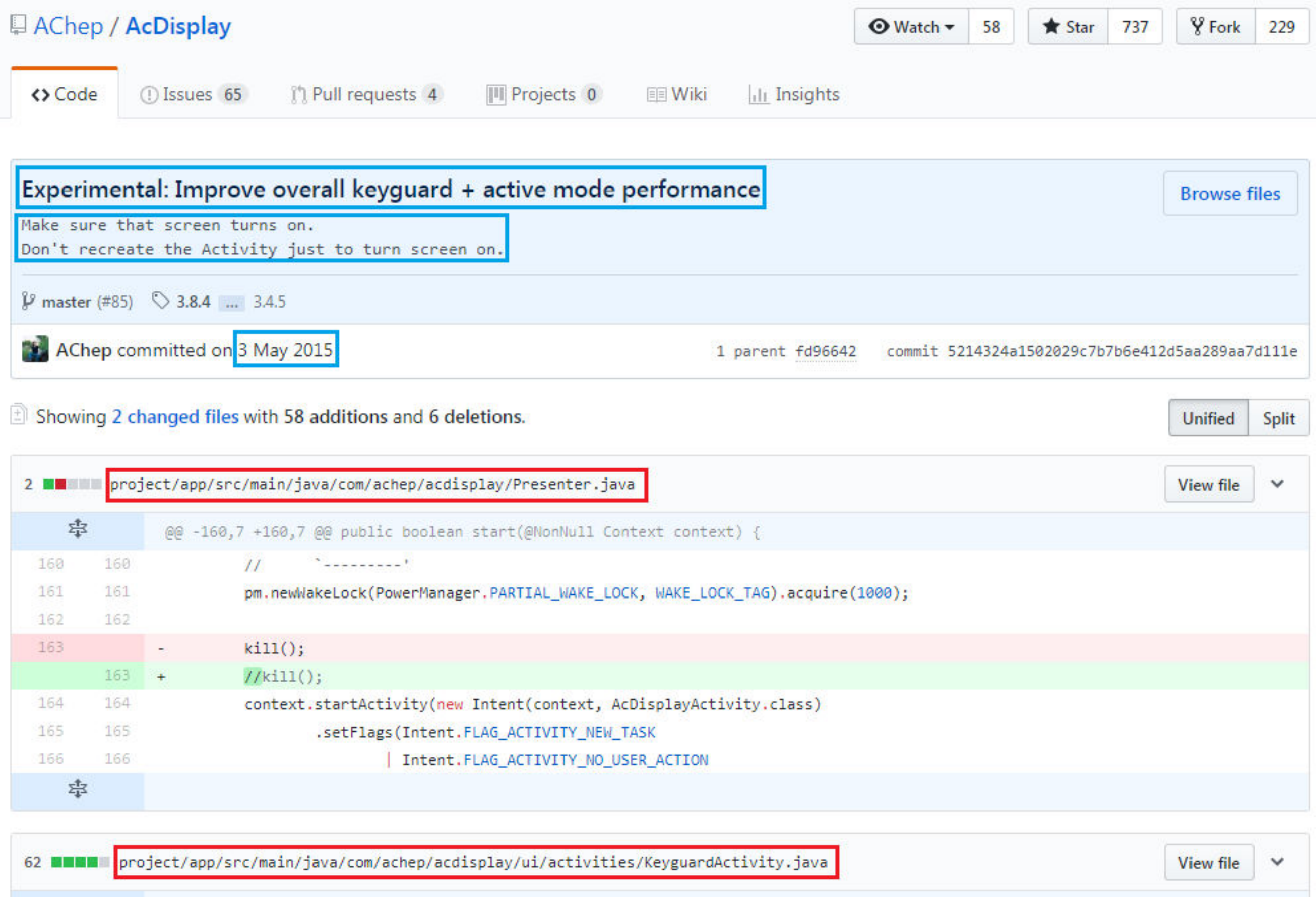}
	\vspace{-3mm}
	\caption{Commit Message Illustration} \label{fig:commit}
\end{figure}

Fig.~\ref{fig:commit} shows a commit example from AcDisplay. We extract title, description, timestamps (in blue rectangle) and relevant file paths (in red rectangle) information. All the files will be tagged with such information. Particularly, in this step we only consider the source code files and their related commit messages. The irrelevant commits (e.g., those do not involve source code files changes which are usually associated with  `.html', `.properties', or `.md' files) are removed in the first place. 

\subsubsection{Localization}

\noindent\textbf{Similarity Computation}.
As mentioned earlier, due to the semantic gap between natural language and programming language, the direct similarity matching cannot precisely localize potential change files. We introduce the commit information to bridge the gap. Therefore, the similarity is attributed to the following two parts:
\begin{itemize}
	\item the similarity between the user review and the code components extracted from one class of the target app; 
	
	\item the similarity between the user review and the commit tags of one class whose time stamps were earlier than the user review.
\end{itemize}

Palomba et al. \cite{palomba15} used the asymmetric Dice coefficient \cite{baeza1999modern} to compute a textual similarity between a user review and a commit, as well as a textual similarity between a user review and an issue. Since user reviews are usually much shorter than source code files and commits, asymmetric Dice coefficient based similarity measures are usually employed (as opposed to other alternatives such as the cosine similarity or the Jaccard coefficient~\cite{jaccard1901etude}). However, the original asymmetric Dice coefficient treats all the word equally and ignores those words which occur more frequently. Hence, we introduce a weighted asymmetric Dice coefficient as follows: 
\begin{equation} \label{eq:sim1}
sim(r_i,code_j) = \frac{\sum\limits_{w_k\in W_{r_i}\cap W_{code_j}} df_{w_k} }{\min \left(\sum\limits_{w_r\in W_{r_i}}df_{w_r}, \sum\limits_{w_c\in W_{code_j}}df_{w_c}\right)}
\end{equation}
where $W_{r_i}$ is the set of words within the review $r_i$, $W_{code_j}$ is the set of words within the code components of class $j$, $df_{w_k}$ represents the document frequency (df) of the word $w_k$,
and the $\min(\cdot,\cdot)$ function returns the argument whose value is smaller.  
In \eqref{eq:sim1}, we use $df$'s value as the weight of the words. 
The intuition is that the more frequently a word occurs, the more important the word is. 

The similarity between a user review and commit tags is computed analogously, by replacing $W_{code_j}$ by $W_{commit_j}$ as shown in \eqref{eq:sim2}, where $W_{commit_j}$ is the set of words within the commit tags of class $j$.

\begin{equation} \label{eq:sim2}
sim(r_i,commit_j) = \frac{\sum\limits_{w_k\in W_{r_i}\cap W_{commit_j}} df_{w_k} }{\min\left(\sum\limits_{w_r\in W_{r_i}}df_{w_r}, \sum\limits_{w_c\in W_{commit_j}}df_{w_c}\right)}
\end{equation}

\medskip
\noindent\textbf{Dynamic Interpolation Weights}.
The similarity score between user reviews and source code files is calculated by a linear combination of the similarity score between the reviews and the source code contained in the files and the one between the reviews and the commit messages associated with the files (cf.\ Section~\ref{sect:tag}).  
However, in the initial stage of the project life cycle, there is no enough commit information, a reminiscent of the cold-start problem. 
During the course of the project, commit messages accumulate. In light of this,   
we dynamically assign the weights to the two parts, inspired by dynamic interpolation weights~\cite{tu14,knight09}:
\[
  sim(r_i,file_j) = \frac{L-\gamma}{L}sim(r_i,code_j) + \frac{\gamma}{L}sim(r_i,commit_j)
\]
where $\gamma$ is the number of common words which appear in both user review $r_i$ and commit tags $commit_j$, and $L$ is the number of words in user review $r_i$. We use $L$ instead of the concentration parameter because we can determine the maximum number of $\gamma$. Based on the above equation, if $class_j$ does not have enough commit tags (when $\gamma$ is small), then the code components of $class_j$ will be preferred, which can cope with the cold-start problem in which there are few commits or even no commits at the beginning of project life cycle. As the number of commit tags is growing, when $\gamma$ is large, the commits will be preferred. This strategy could gradually increase the weight of commit messages during similarity calculation over time.


\section{Case Study}\label{case}
We collect 
the user reviews and commit messages of ten popular apps available from Google Play. The basic statistics of the selected projects are listed in Table~\ref{tab:proj}.
\begin{table}[ht]
	\caption{Overview of selected apps}
	\vspace{-3mm}
	\centering
	\scalebox{0.80}{
		\begin{tabular}{|c|c|c|c|c|}
			\hline
			App Name & Category & Version & Comm. Msg. No. & Review No.\\
			\hline\hline
			AcDisplay &  Personalization & 3.8.4 & 1096 & 8074 \\
			SMS Backup+ & Tools & 1.5.11 & 1687 & 1040 \\
			AnySoftKeyboard & Tools & 1.9 & 4227 & 3043 \\
			Phonograph & Music\&Audio & 1.2.0 & 1470 & 6986 \\
			Terminal Emulator & Tools & 1.0.70 & 1030 & 4351 \\
			SeriesGuide & Entertainment & 44 & 9217 & 5287 \\
			ConnectBot & Communication & 1.9.5 & 1714 & 4806 \\
			Signal & Communication & 4.29.5 & 3846 & 6460 \\
			AntennaPod & Video Players & 1.7.0 & 4562 & 3708 \\
			K-9 Mail & Communication & 5.600 & 8005 & 8040 \\\hline\hline
			Total & - & - & 36854 & 58775 \\\hline
		\end{tabular}
		\vspace{-3mm}
		\label{tab:proj}
	}
\end{table}
The selection criteria for Android apps are (i) open-source Android apps published on the Google Play market with version system and commit messages publicly accessible, and (ii) diversity in terms of app category (e.g., Personalization, Tools, Communication), size, and number of commit messages.

We developed two web scrapers to crawl the raw data: 
one is to extract user reviews from Google Play Store, and the other is to extract commit messages from GitHub. As for the source code, we download the latest version of the apps from GitHub. The version information is also shown in the table.

To evaluate how well our approach could help developers localize potential change files, we 
investigate the following two research questions.
\begin{description}
	\item[RQ1:] Does the constraint-based clustering algorithm perform better?
	\item[RQ2:] Do commit messages improve the accuracy of localization?
\end{description}


\medskip
\noindent{\textbf{RQ1}.}
We implemented the approach and ran it on our dataset to address the above research questions. We chose ChangeAdvisor~\cite{palomba17} as the baseline for two reasons. Firstly, ChangeAdvisor is the  closest/most relevant approach with ours and the two address the same question largely; secondly, ChangeAdvisor is the 
state-of-the-art reference on clustering of user reviews in  literature, and its superiority had been demonstrated compared to other similar approaches, such as BLUiR~\cite{Saha:ase13}.  We observe that in the previous work~\cite{palomba17}, the authors did not distinguish the two general categories of user reviews, i.e., feature request and problem discovery, in the clustering process. Thus it is very likely that reviews of the two categories are clustered together. For example, in AcDisplay, the two reviews ``The error was with my phone and other apps were crashing as well'' and ``It lacks UI customization like resizing the clock size or adding circle boarder ...'' are grouped into the same cluster. Apparently, they express quite different concerns. To give a fair comparison, we differentiate the two categories, reuse the original prototype of ChangeAdvisor and the same parameter settings as published in related work~\cite{palomba17}.

\begin{table}[ht]
	\caption{Apps' review information (FR: feature request; PD: problem discovery)}
	\vspace{-3mm}
	\centering
	\scalebox{0.78}{
		\begin{tabular}{|c|c|c|c|c|c|c|c|}
			\hline
			\multirow{2}{*}{App Name} & Review& \multicolumn{3}{c|}{FR No. } & \multicolumn{3}{c|}{PD No.}\\\cline{3-8}
			& No. & Total & M-link & N-link & Total & M-link & N-link\\
			\hline\hline
			AcDisplay & 8074 & 1400 & 50 & 50 & 1437 & 50 & 50\\
			SMS Backup+ & 1040 & 677 & 22 & 8 & 1425 & 32 & 21\\
			AnySoftKeyboard & 3043 & 280 & 25 & 3 & 290 & 16 & 6\\
			Phonograph & 6986 & 1094 & 63 & 28 & 960 & 42 & 53\\
			Terminal Emulator & 4351 & 248  & 13 & 9 & 372 & 10 & 28 \\
			SeriesGuide & 5287 & 588 & 28 & 21 & 460 & 16 & 21\\
			ConnectBot  & 4806 & 467 & 29 & 16 &  604 & 43 & 17\\
			Signal & 6460 & 629 & 36 & 23 & 792 & 32 & 45 \\
			AntennaPod & 3708 & 336 & 25 & 21 & 359 & 16 & 22\\
			K-9 Mail & 8040 & 1018 & 65 & 36 & 1854 & 66 & 28 \\\hline\hline
			Total & 58775 & 6737  & 356 & 215 & 8553 & 323 & 291 \\\hline
		\end{tabular}
		\vspace{-3mm}
		\label{tab:review}
	}
\end{table}

To annotate the constraints, we asked three researchers in software engineering (including the second author), to randomly select a small amount of reviews as samples from the whole dataset. These samples are inspected independently and the resulting constraints are collectively validated. Table~\ref{tab:review} shows the information of the input to the clustering stage. As mentioned before, we distinguish two categories, i.e., feature request (FR) and problem discovery (PD). The table also gives the human-annotated cluster constraints information of each app. In total, in the feature request category, the annotated must-link (M-link) takes up around 10.57\% (356 pairs out of 6,737), and cannot-link (N-link) takes up around 6.38\% (215 pairs out of 6,737); while in the problem discovery category, the percentage of must-link instances is around 7.55\% (323 pairs out of 8,553), and of cannot-link is around 6.80\% (291 pairs out of 8553). The marked instances are randomly selected from each app.  In line with the metrics used in the baseline work~\cite{palomba17}, we compare the \emph{cohesiveness} and \emph{separation} of clustering results between two approaches.

RISING incorporates domain knowledge to annotate the must-link and cannot-link information to the subset of user reviews, and leverages a heuristic-based method to infer the hyperparameter $K$. ChangeAdvisor directly applies the Hierarchical Dirichlet Process (HDP) algorithm to group the sentences of the reviews~\cite{teh05}. So we first need to compare the cluster numbers that the two approaches yield, and then the quality of each cluster.

\begin{table}[ht]
	\caption{Clustering information}
	\vspace{-3mm}
	\centering
	\scalebox{1}{
		\begin{tabular}{|c|c|c|c|c|}
			\hline
			\multirow{2}{*}{App Name} & \multicolumn{2}{c|}{ChangeAdvisor} & \multicolumn{2}{c|}{RISING} \\\cline{2-5}
			& FR No. & PD No. & FR No. & PD No. \\
			\hline\hline
			AcDisplay & 10 & 12 & 83 & 129\\
			SMS Backup+ & 12 & 10 & 71 & 107\\
			AnySoftKeyboard & 11 & 11 & 46 & 41\\
			Phonograph & 8 & 7 & 105 & 106\\
			Terminal Emulator & 11 & 8 & 38 & 45\\
			SeriesGuide & 11 & 11 & 59 & 44\\
			connectBot  & 8 & 11 & 47 & 75\\
			Signal & 12 & 10 & 85 & 97\\
			AntennaPod & 10 & 8 & 43 & 57\\
			K-9 Mail & 6 & 10 & 113 & 154\\\hline\hline
			Total & 99 & 98 & 690 & 855\\\hline
		\end{tabular}
		\vspace{-3mm}
		\label{tab:cluster}
	}
\end{table}

Table~\ref{tab:cluster} presents the comparison of clustering between our approach and ChangeAdvisor. From the table, we can observe that, the number of clusters yielded by ChangeAdvisor and our approach varies a lot. 
The median cluster values of feature request and problem discovery categories in the studied apps by ChangeAdvisor are 10.5 and 10, respectively. However, in our approach, the median cluster values of the two categories are 65 and 86, respectively. Moreover, we found that, the clustering result is quite unbalanced in ChangeAdvisor. For example, in AcDisplay, the number of clusters of ChangeAdvisor (Problem Discovery Category) is 12, but 1142 out of 1437 sentences are grouped into one cluster, which takes up more than 79\%; while in our approach, the largest cluster contains 339 instance sentences, which takes up around 23.6\%. This highlights that the clusters generated by our approach are of more practical use to developers compared to the one generated by ChangeAdvisor.

To compare the quality of clusters obtained by two approaches, we use the Davis-Bouldin index (DBI), a widely adopted method to evaluate clustering algorithms~\cite{4766909}, as a metric to assess the cohesiveness of intra-clusters and the separation of inter-clusters. This is an internal evaluation scheme, where the validation of how well the clustering has been done is made using quantities and features inherent to the dataset. DBI consists of two parts, one is the measure of scatter within the cluster, and the other is of separation between clusters.

For a cluster $C=\{X_1, \cdots, X_T\}$, the measure of scatter $S_C$ of $C$ is defined as follows.
\begin{equation}
 S_C = \left(\frac{1}{T}\sum_{j=1}^{T}\left| X_{j} - A \right|^{2} \right)^{1/2}
\end{equation}
where  $T$ is the size of the cluster and
$A$ is the centroid of the cluster $C$.

The measure $M_{i,j}$ of separation between clusters $C_i$ and $C_j$ is defined as follows.
\begin{equation}
M_{ i , j } = \left\| A _ { i } - A _ { j } \right\| _ { 2 } = \left( \sum _ { k = 1 } ^ { n } \left| a _ { k , i } - a _ { k , j } \right| ^ { 2 } \right) ^ { \frac { 1 } { 2 } }
\end{equation}
where $a_{k , i }$ is the $k^{th}$ element of the centriod $A_{i}$ of cluster $C_i$.
DBI can then be defined as
\begin{equation}
 DBI := \frac { 1 } { K } \sum _ { i = 1 } ^ { K } \max _ { j \neq i } \frac { S _ { i } + S _ { j } } { M _ { i , j } }
\end{equation}
where $K$ is the number of clusters.

The DBI value is a standard measurement of the quality of clustering, i.e., the cohesiveness of intra-cluster and the separation of inter-clusters. Typically, a better clustering algorithm 
admits a \emph{lower} value of DBI.

ChangeAdvisor uses HDP algorithm for clustering, which accepts text objects as input. 
To enable the adoption of DBI, we need to convert the dataset of ChangeAdvisor into a vector format. In order to ensure a fair comparison, we use the same method  to convert the transformed sentences into the vector representation as in our approach, as detailed in Section~\ref{sec:cluster}.


\begin{table}[ht]
	\caption{DBI results comparison}
	\vspace{-3mm}
	\centering
	\scalebox{1}{
		\begin{tabular}{|c|c|c|c|c|}
			\hline
			\multirow{2}{*}{App Name} & \multicolumn{2}{c|}{ChangeAdvisor} & \multicolumn{2}{c|}{RISING} \\\cline{2-5}
			& FR & PD & FR & PD \\
			\hline\hline
			AcDisplay & 0.493 & 0.361 & 0.035 & 0.020\\
			SMS Backup+ & 0.321 & 0.444 & 0.047 & 0.042\\
			AnySoftKeyboard & 0.357 & 0.342 & 0.050 & 0.050\\
			Phonograph & 0.514 & 0.693 & 0.031 & 0.029\\
			Terminal Emulator & 0.300 & 0.557 & 0.105 & 0.060\\
			SeriesGuide & 0.440 & 0.303 & 0.075 & 0.057\\
			ConnectBot  & 0.606 & 0.479 & 0.080 & 0.027\\
			Signal & 0.317 & 0.391 & 0.055 & 0.027\\
			AntennaPod & 0.447 & 0.548 & 0.048 & 0.046\\
			K-9 Mail & 0.928 & 0.538 & 0.040 & 0.022\\\hline\hline
			Average & 0.472 & 0.466 & 0.057 & 0.038\\\hline
		\end{tabular}
		\vspace{-3mm}
		\label{tab:dbi}
	}
\end{table}

The results are summarized in Table~\ref{tab:dbi}. From the table, we can observe that  our approach yields a considerably better DBI result compared with ChangeAdvisor. The average DBI values of feature request and problem discovery by ChangeAdvisor  are 0.472 and 0.466 respectively; while by our approach, the average values are 0.057 and 0.038 respectively.

In addition,  to demonstrate the effectiveness of  review segmentation, we conduct an ablation study for RISING and RISING without review segmentation. The effectiveness is also measured by DBI values. The results are shown in Table~\ref{tab:dbireviewsegmentation} which suggest that the review segmentation operation has indeed improved the clustering effect by reducing the average DBI values of feature request and problem discovery from 0.167 to 0.057 and 0.159 to 0.038, respectively.

\begin{table}[ht]
	\centering
	\caption{DBI results comparison for the review segmentation}
	\vspace{-3mm}
	\scalebox{1}{
		\begin{tabular}{|c|c|c|c|c|}
			\hline
			\multirow{2}{*}{App Name} & \multicolumn{2}{c|}{RISING (no segmentation)} & \multicolumn{2}{c|}{RISING} \\\cline{2-5}
			& FR & PD & FR & PD \\
			\hline\hline
			AcDisplay & 0.235 & 0.247 & 0.035 & 0.020\\
			SMS Backup+ & 0.094 & 0.110 & 0.047 & 0.042\\
			AnySoftKeyboard & 0.159 & 0.184 & 0.050 & 0.050\\
			Phonograph & 0.178 & 0.131 & 0.031 & 0.029\\
			Terminal Emulator & 0.156 & 0.120 & 0.105 & 0.060\\
			SeriesGuide & 0.159 & 0.125 & 0.075 & 0.057\\
			ConnectBot  & 0.250 & 0.230 & 0.080 & 0.027\\
			Signal & 0.129 & 0.113 & 0.055 & 0.027\\
			AntennaPod & 0.151 & 0.164 & 0.048 & 0.046\\
			K-9 Mail & 0.159 & 0.165 & 0.040 & 0.022\\\hline\hline
			Average & 0.167 & 0.159 & 0.057 & 0.038\\\hline
		\end{tabular}
		\vspace{-3mm}
		\label{tab:dbireviewsegmentation}}
\end{table}

To further evaluate the quality of clustered reviews, we hired mobile  app developers as external evaluators. Specifically, we contacted over 10 professional mobile developers from our personal network applying the following participant constraints: they were not familiar with the actual apps under study and they had at least three years of mobile development experience. This selection process allowed us to receive feedback by only external participants having the adequate experience to the context of the tasks. Considering also the developers' availability and willingness to perform the study, we hired in total five developers, three to evaluate the quality of clustered reviews, while all of them participated in the evaluation tasks of RQ2 (as reported later in the paper). Thus, we asked the three of the mobile developers to look into the clustered sentences and to assess the coherence of content of each individual cluster as well as the semantics separation of different clusters. The assessment is given in Likert scale grades: ``exactly related topics'' (5), ''mostly related topics'' (4), ``basically related topics'' (3), ``just a few related topics'' (2), and ``not relevant topics'' (1). Different from the evaluation method in the baseline work~\cite{palomba17}, we evaluate all the clusters, and calculate the average value as the final result.

The results are shown in Table~\ref{tab:likert}. From the table, we observe that RISING yields a better value of Likert scale compared with ChangeAdvisor. The average values of feature request and problem discovery categories by ChangeAdvisor are 2.07 and 1.94 respectively; while by RISING, the average values are 4.20 and 4.26 respectively.

\begin{table}[ht]
	\caption{Likert results comparison}
	\vspace{-3mm}
	\centering
	\scalebox{1}{
		\begin{tabular}{|c|c|c|c|c|}
			 \hline
			 \multirow{2}{*}{App Name} & \multicolumn{2}{c|}{ChangeAdvisor} & \multicolumn{2}{c|}{RISING} \\\cline{2-5}
			 & FR & PD & FR & PD \\
			 \hline\hline
			 AcDisplay & 2.22 & 2.12 & 4.30 & 4.29\\
			 SMS Backup+ & 1.93 & 2.03 & 4.23 & 4.26\\
			 AnySoftKeyboard & 2.50 & 2.47 & 4.23 & 4.09\\
			 Phonograph & 2.35 & 1.55 & 4.40 & 4.35\\
			 Terminal Emulator & 2.18 & 2.15 & 3.83 & 4.17\\
			 SeriesGuide & 2.17 & 1.74 & 4.22 & 4.29\\
			 ConnectBot & 1.43 & 2.05 & 4.20 & 4.35\\
			 Signal & 1.96 & 1.70 & 4.26 & 4.31\\
			 AntennaPod & 2.08 & 1.67 & 4.17 & 4.25\\
			 K-9 Mail & 1.87 & 1.92 & 4.11 & 4.25\\\hline\hline
			 Average & 2.07 & 1.94 & 4.20 & 4.26\\\hline
		\end{tabular}
		\vspace{-3mm}
		\label{tab:likert}
	}
\end{table}
\textit{The above objective and subjective measures answer RQ1 that our constraints-based clustering method, aided by more intensive (automated) data preprocessing and marginal human annotation efforts, could greatly boost the clustering performance.}


\medskip
\noindent{\textbf{RQ2}.} 
To address RQ2, we need to judge whether commit messages improve the accuracy of localization. In the experiments, we use the same ten Android apps in the preceding step (cf.~Table~\ref{tab:proj}). As the first step, we need to obtain the ground-truth for the localization result which requires human examination. As mentioned in the design of RQ1, to reduce personal bias,  we hired two additional mobile app developers, both of whom have over three years of development experience as well.
The five evaluators were asked to check the localization result individually. Only after the individual evaluation, they discussed jointly trying to reach a consensus. The discussion happened via conference calls involving all the five evaluators (as physical meetings turned out to be difficult to organize). During the meeting, the evaluator of the work led the discussion.
	All participants quickly separated 
	the cases in which there was a clear agreement from those in which no agreement can be quickly drawn. 
	Extensive discussions then focused on those more difficult cases 
	until a consensus was reached.  Results of this evaluation process serves as the ground-truth of RQ2.

As the next step, we apply ChangeAdvisor  and RISING to the reviews to compare the results returned from them against the ground-truth results from the previous step.
For each category in each app, we randomly select 12-18 user reviews and then apply ChangeAdvisor and RISING separately to these sample reviews. (In ChangeAdvisor, a sample size of around 10 for each app category was used. We generally follow the guideline of the baseline.) Overall, we select 297 (150 + 147) user reviews from these 10 apps. RISING could return potential change files in all the cases. However, ChangeAdvisor could only give outputs when inputting 132 (82 + 50) user reviews,  less than 50\% of RISING. The details of the localizability comparison of the two approaches in the studied apps are given in Table~\ref{tab:localize}.

To evaluate the localization result, we employed the Top-k accuracy and NDCG as metrics which are commonly used in recommendation systems~\cite{DBLP:conf/issre/TantithamthavornTIM13, DBLP:conf/snpd/TantithamthavornIM13,DBLP:journals/corr/abs-1810-09723}. Top-k accuracy can be calculated as
\[
\operatorname {Top-k} accuracy( U ) = \frac { \sum\limits_{ u \in U } \operatorname { isCorrect } ( u , \operatorname { Top-k} ) } { | U | }
\]
where $U$ represents the set of all user feedbacks and the $\operatorname { isCorrect } ( r , \operatorname { Top-k} )$ function returns 1 if at least one of $\operatorname {Top-k}$ source code files actually is relevant to the user feedback $u$; and returns 0 otherwise.

Table~\ref{tab:localize_accuracy} reports the Top-k accuracy of ChangeAdvisor and RISING for each of the considered apps where the value $k$ is set to be 1, 3 and 5.  From the table, we can observe that,  in most cases, RISING substantially outperforms ChangeAdvisor in terms of Top-k hitting. On average, for feature request category, the Top-1, Top-3, Top-5 values can be improved from 44.64\% to 76.74\%, 70.54\% to 91.77\%, and 76.65\% to 98.00\% respectively; for problem discovery category, the Top-1, Top-3, Top-5 values are improved from 48.50\% to 76.04\%, 65.08\% to 93.84\%, and 76.00\% to 98.04\% respectively.

NDCG is defined as follows:
\[
N D C G @ k = \frac { D C G @ k } { I D C G @ k }, \quad \left( D C G @ k = \sum _ { i = 1 } ^ { k } \frac { r _ { i } } { \log_2 (i + 1) } \right)
\]
where $r_{i}$ = 1 if the $i$-th source code file is related to the user feedback, and $r _ { i }$ = 0 otherwise. IDCG is the ideal result of DCG, which means all related source code files are ranked higher than the unrelated ones. For example, if an algorithm recommends five source code files in which the 1-st, 3-rd and 5-th source code files are related, the results are represented as $\{1,0,1,0,1\}$, whereas the ideal result is $\{1,1,1,0,0\}$.

Table~\ref{tab:localize_ndcg} reports the NDCG values of ChangeAdvisor and RISING for each of the considered apps where, similarly, the value $k$ is set to be 1, 3 and 5. Based on the table, we observe that, in most cases of the studied apps, the NDCG value of RISING is greater than that of ChangeAdvisor, which indicates a better performance. On average, the NDCG@1, NDCG@3 and NDCG@5 values of ChangeAdvisor in the problem discovery category are 48.50\%, 46.37\%, and 59.69\% respectively. In contrast, the corresponding values of RISING in this category are 76.03\%, 74.46\%, and 85.95\% respectively. In feature request category, the NDCG@1, NDCG@3 and NDCG@5 values of ChangeAdvisor are 44.64\%, 53.52\%, 60.94\% respectively; while the values of RISING in this category are 76.74\%, 74.11\%, and 85.79\% respectively.


To further demonstrate the impact of commit messages for localization, we localize the user reviews by merely using source code information. The localization results are shown in Table~\ref{tab:localize_sourcecodeonly}. By comparing the localization results between RISING (Source Code) and RISING, we can observe that, for the feature request category, the Top-1, Top-3, Top-5 values are improved from 67.53\%, 83.91\% and 91.48\% to 76.74\%, 91.77\% and 98.00\%, and the NDCG@1, NDCG@3, NDCG@5 are improved from 67.53\%, 65.96\%, 78.09\% to 76.74\%, 74.11\%, 85.79\%; for the problem discovery category, the Top-1, Top-3, Top-5 values are improved from 63.67\%, 85.52\%, 92.46\% to 76.04\%, 93.84\%, 98.04\%, and the NDCG@1, NDCG@3, NDCG@5 are improved from 63.67\%, 64.76\%, 77.20\% to 76.03\%, 74.46\%, 85.95\%. The results demonstrate that  commit messages do contribute to improving the localization performance.

\emph{The experiment results answer RQ2 that, in terms of the localization accuracy, our approach which exploits commit messages to fill the lexicon gap could  improve the performance greatly.}

\begin{table}[ht]
	\caption{Overview of Localization}
	\vspace{-3mm}
	\centering
	\scalebox{1}{
		\begin{tabular}{|c|c|c|c|c|}
			\hline
			\multirow{2}{*}{App Name} & \multicolumn{2}{c|}{ChangeAdvisor} & \multicolumn{2}{c|}{RISING} \\\cline{2-5}
			& FR No. & PD No. & FR No. & PD No. \\
			\hline\hline
			AcDisplay & 6 & 2 & 17 & 15\\
			SMS Backup+ & 7 & 2 & 14 & 15\\
			AnySoftKeyboard & 7 & 5 & 14 & 13\\
			Phonograph & 9 & 4 & 18 & 16\\
			Terminal Emulator & 4 & 6 & 15 & 16\\
			SeriesGuide & 14 & 10 & 15 & 14\\
			ConnectBot & 10 & 4 & 14 & 16\\
			Signal & 7 & 8 & 16 & 14\\
			AntennaPod & 10 & 5 & 15 & 14\\
			K-9 Mail & 8 & 4 & 12 & 14\\\hline\hline
			Sum & 82 & 50 & 150 & 147\\\hline
		\end{tabular}
		\vspace{-3mm}
		\label{tab:localize}
    }
\end{table}

\begin{table*}[ht]
 	\caption{Top-k Accuracy of Localization}
	\vspace{-3mm}
	\centering
	\scalebox{1.04}{
		\begin{tabular}{|c|c|c|c|c|c|c|c|c|c|c|c|c|}
			\hline
			\multirow{3}{*}{App Name} & \multicolumn{6}{c|}{ChangeAdvisor} & \multicolumn{6}{c|}{RISING} \\\cline{2-13}
            & \multicolumn{3}{c|}{FR} &\multicolumn{3}{c|}{PD}& \multicolumn{3}{c|}{FR} &\multicolumn{3}{c|}{PD}\\\cline{2-13}
			& Top-1 & Top-3 & Top-5 & Top-1 & Top-3 & Top-5& Top-1 & Top-3 & Top-5 & Top-1 & Top-3 & Top-5\\
			\hline\hline
			AcDisplay & 0.8333 & 0.8333 & 0.8333 & 0.5000 & 0.5000 & 0.5000 & 1.0000 & 1.0000 & 1.0000 & 0.8000 & 1.0000 & 1.0000\\
			SMS Backup+ & 0.2857 & 0.5714 & 0.7143 & 0.5000 & 0.5000 & 0.5000 & 0.7143 & 1.0000 & 1.0000 & 0.8000 & 1.0000 & 1.0000\\
			AnySoftKeyboard & 0.7143 & 0.8571 & 0.8571 & 0.8000 & 1.0000 & 1.0000 & 1.0000 & 1.0000 & 1.0000 & 0.9231 & 1.0000 & 1.0000\\
			Phonograph & 0.6667 & 0.7778 & 0.7778 & 0.5000 & 0.7500 & 1.0000 & 0.8333 & 1.0000 & 1.0000 & 0.9375 & 0.9375 & 1.0000\\
			Terminal Emulator & 0.2500 & 0.7500 & 0.7500 & 0.5000 & 0.8333 & 1.0000 & 0.8667 & 0.9333 & 0.9333 & 0.6875 & 0.9375 & 0.9375\\
			SeriesGuide & 0.2857 & 0.6429 & 0.7857 & 0.4000 & 0.7000 & 0.8000 & 0.6000 & 0.8000 & 0.8667 & 0.7143 & 1.0000 & 1.0000\\
			ConnectBot & 0.3000 & 0.7000 & 0.7000 & 0.7500 & 0.7500 & 0.7500 & 0.6429 & 0.7857 & 1.0000 & 0.8125 & 0.9375 & 0.9375\\
			Signal & 0.4286 & 0.5714 & 0.5714 & 0.2500 & 0.3750 & 0.7500 & 0.5000 & 0.8750 & 1.0000 & 0.6429 & 0.8571 & 1.0000\\
			AntennaPod & 0.2000 & 0.6000 & 0.8000 & 0.4000 & 0.6000 & 0.8000 & 0.6000 & 0.8667 & 1.0000 & 0.7857 & 0.9286 & 0.9286\\
			K-9 Mail & 0.5000 & 0.7500 & 0.8750 & 0.2500 & 0.5000 & 0.5000 & 0.9167 & 0.9167 & 1.0000 & 0.5000 & 0.7857 & 1.0000\\\hline\hline
			Average & 0.4464 & 0.7054 & 0.7665 & 0.4850 & 0.6508 & 0.7600 & 0.7674 & 0.9177 & 0.9800 & 0.7604 & 0.9384 & 0.9804\\\hline
		\end{tabular}
		\vspace{-3mm}
		\label{tab:localize_accuracy}
	}
\end{table*}

\begin{table*}[ht]
	\caption{NDCG@k of Localization}
	\vspace{-3mm}
	\centering
	\scalebox{0.80}{
		\begin{tabular}{|c|c|c|c|c|c|c|c|c|c|c|c|c|}
			\hline
			\multirow{3}{*}{App Name} & \multicolumn{6}{c|}{ChangeAdvisor} & \multicolumn{6}{c|}{RISING} \\\cline{2-13}
            & \multicolumn{3}{c|}{FR} &\multicolumn{3}{c|}{PD}& \multicolumn{3}{c|}{FR} &\multicolumn{3}{c|}{PD}\\\cline{2-13}
			& NDCG@1 & NDCG@3 & NDCG@5 & NDCG@1 & NDCG@3 & NDCG@5 & NDCG@1 & NDCG@3 & NDCG@5 & NDCG@1 & NDCG@3 & NDCG@5\\
			\hline\hline
			AcDisplay & 0.8333 & 0.7675 & 0.8012 & 0.5000 & 0.3520 & 0.4427 & 1.0000 & 0.8843 & 0.9565 & 0.8000 & 0.7996 & 0.8973\\
			SMS Backup+ & 0.2857 & 0.3733 & 0.4726 & 0.5000 & 0.3066 & 0.4386 & 0.7143 & 0.6598 & 0.8180 & 0.8000 & 0.6837 & 0.8706\\
			AnySoftKeyboard & 0.7143 & 0.7672 & 0.8000 & 0.8000 & 0.6922 & 0.8692 & 1.0000 & 0.8630 & 0.9611 & 0.9231 & 0.8989 & 0.9507\\
			Phonograph & 0.6667 & 0.6169 & 0.7021 & 0.5000 & 0.5180 & 0.7303 & 0.8333 & 0.8010 & 0.9120 & 0.9375 & 0.7896 & 0.9242\\
			Terminal Emulator & 0.2500 & 0.5044 & 0.5705 & 0.5000 & 0.6087 & 0.7509 & 0.8667 & 0.7994 & 0.8872 & 0.6875 & 0.6753 & 0.8127\\
			SeriesGuide & 0.2857 & 0.4422 & 0.5443 & 0.4000 & 0.3981 & 0.5572 & 0.6000 & 0.6013 & 0.7328 & 0.7143 & 0.7538 & 0.8666\\
			ConnectBot & 0.3000 & 0.4429 & 0.5279 & 0.7500 & 0.7500 & 0.7500 & 0.6429 & 0.6456 & 0.7962 & 0.8125 & 0.8265 & 0.8877\\
			Signal & 0.4286 & 0.4885 & 0.4885 & 0.2500 & 0.2984 & 0.4874 & 0.5000 & 0.6676 & 0.7693 & 0.6429 & 0.7113 & 0.8188\\
			AntennaPod & 0.2000 & 0.4028 & 0.5406 & 0.4000 & 0.3860 & 0.5564 & 0.6000 & 0.6756 & 0.8246 & 0.7857 & 0.7262 & 0.8232\\
			K-9 Mail & 0.5000 & 0.5460 & 0.6467 & 0.2500 & 0.3266 & 0.3859 & 0.9167 & 0.8131 & 0.9208 & 0.5000 & 0.5808 & 0.7430\\\hline\hline
			Average & 0.4464 & 0.5352 & 0.6094 & 0.4850 & 0.4637 & 0.5969 & 0.7674 & 0.7411 & 0.8579 & 0.7603 & 0.7446 & 0.8595\\\hline
		\end{tabular}
		\vspace{-3mm}
		\label{tab:localize_ndcg}
	}
\end{table*}

\begin{table*}[ht]
 	\caption{Localization of RISING (Source Code Only)}
	\vspace{-3mm}
	\centering
	\scalebox{0.91}{
		\begin{tabular}{|c|c|c|c|c|c|c|c|c|c|c|c|c|}
			\hline
			\multirow{2}{*}{App Name} & \multicolumn{12}{c|}{RISING(sourceCodeOnly)}\\\cline{2-13}
            & \multicolumn{3}{c|}{FR} &\multicolumn{3}{c|}{PD}& \multicolumn{3}{c|}{FR} &\multicolumn{3}{c|}{PD}\\\cline{2-13}
			& Top-1 & Top-3 & Top-5 & Top-1 & Top-3 & Top-5 & NDCG@1 & NDCG@3 & NDCG@5 & NDCG@1 & NDCG@3 & NDCG@5\\
			\hline\hline
			AcDisplay & 0.9412 & 1.0000 & 1.0000 & 0.7333 & 0.9333 & 1.0000 & 0.9412 & 0.8631 & 0.9394 & 0.7333 & 0.7124 & 0.8547\\
			SMS Backup+ & 0.5714 & 1.0000 & 1.0000 & 0.6000 & 0.9333 & 0.9333 & 0.5714 & 0.6567 & 0.7974 & 0.6000 & 0.6387 & 0.7676\\
			AnySoftKeyboard & 0.9286 & 0.9286 & 1.0000 & 0.8462 & 0.8462 & 1.0000 & 0.9286 & 0.7975 & 0.9189 & 0.8462 & 0.7396 & 0.8717\\
			Phonograph & 0.7778 & 0.8889 & 0.9444 & 0.8125 & 0.9375 & 1.0000 & 0.7778 & 0.7009 & 0.8359 & 0.8125 & 0.7591 & 0.8807\\
			Terminal Emulator & 0.6667 & 0.8000 & 0.8667 & 0.6875 & 0.8125 & 0.8750 & 0.6667 & 0.6536 & 0.7604 & 0.6875 & 0.6658 & 0.7618\\
			SeriesGuide & 0.4667 & 0.6000 & 0.7333 & 0.4286 & 0.8571 & 0.9286 & 0.4667 & 0.5020 & 0.5936 & 0.4286 & 0.6107 & 0.7246\\
			ConnectBot & 0.5714 & 0.8571 & 0.9286 & 0.6875 & 0.8750 & 0.9375 & 0.5714 & 0.6202 & 0.7668 & 0.6875 & 0.6666 & 0.7894\\
			Signal & 0.5625 & 0.7500 & 0.8750 & 0.6429 & 0.7857 & 0.8571 & 0.5625 & 0.5224 & 0.6896 & 0.6429 & 0.6154 & 0.7309\\
			AntennaPod & 0.6000 & 0.7333 & 0.8000 & 0.5714 & 0.8571 & 0.9286 & 0.6000 & 0.5532 & 0.6810 & 0.5714 & 0.5815 & 0.7349\\
			K-9 Mail & 0.6667 & 0.8333 & 1.0000 & 0.3571 & 0.7143 & 0.7857 & 0.6667 & 0.7261 & 0.8262 & 0.3571 & 0.4858 & 0.6032\\\hline\hline
			Average & 0.6753 & 0.8391 & 0.9148 & 0.6367 & 0.8552 & 0.9246 & 0.6753 & 0.6596 & 0.7809 & 0.6367 & 0.6476 & 0.7720\\\hline
		\end{tabular}
		\vspace{-3mm}
		\label{tab:localize_sourcecodeonly}
	}
\end{table*}

\section{Discussion}\label{discuss}

Identifying meaningful user reviews from app markets is a non-trivial task, since a majority of them are not informative. Furthermore, to link and localize potential change files based on those meaningful feedbacks would be highly desirable for software developers. Compared with the state-of-the-art baseline work like ChangeAdvisor, RISING could give more fine-grained clustering results and more accurate localization performance. Specifically, after a closer qualitative analysis of the clusters generated by both approaches we found interesting characteristics concerning the clusters generated by RISING, when compared to the one of ChangeAdvisor. First of all, ChangeAdvisor tends to discard a wide majority of reviews, clustering a small subset of informative reviews. For instance, if we consider the app AcDisplay, we observe that the total number of reviews included in all generated clusters, considering both feature requests and problem discovery categories, is 150. This value is drastically smaller than the amount of informative reviews composing the clusters generated by RISING for this app, i.e., 2,053 reviews. As a consequence, the number of clusters for ChangeAdvisor tends to be very small for most projects, compared to our approach, as reported in Table \ref{tab:cluster}. On the other hand, the sizes of the clusters generated by RISING tend to be more balanced, i.e., not too large. Indeed, for RISING, the average number of reviews for each cluster tends to be smaller (11-12 reviews on average v.s. 18-24  for ChangeAdvisor). In our experiments, we also observe that, distinct runs of ChangeAdvisor give noticeably different clustering results, making the clustering less stable or deterministic and less accurate (they tend to be less semantically related).  The clustering result of RISING is much more stable.
To see this, we made two new runs of  RISING and ChangeAdvisor respectively for the app AcDisplay. The results show that the size of the clusters generated by RISING tends to be still very balanced, i.e., not too large, keeping a similar number of clusters, and almost the same average number of reviews for each cluster (11-12 reviews in average).  Vice versa, for ChangeAdvisor we observe for the same app, still very large clusters. Interestingly, in the ChangeAdvisor re-runs, the number of clusters was in one case reduced and in another increased, observing in some cases higher or similar  averages number of reviews for each cluster.
In the localization phase, RISING leverages the commit information to bridge the lexicon gap. Note that commit history contains all the relevant files for the change transaction including not only source files but also configuration related files (such as XML files). Our approach is thus advantageous over other similar approaches to be able to 
locate multiple files which are necessary for problem fix or feature request. In contrast, ChangeAdvisor does not take into account the association between files, which would miss, for instance, configuration files. 

\subsection*{Threats to Validity}
\noindent {\bf Internal validity.}
We conclude that, with domain knowledge, marginal human effort could greatly boost the clustering performance. Such effectiveness has already been demonstrated in various scenarios~\cite{bilenko04,basu08}. In the clustering phase, we only annotate a small portion of the whole review set with must-link and cannot-link, reducing the threat of over-fitting. The recovery of missing traceability links between various software artefacts has also been actively studied in the literature~\cite{Antoniol:tse2002}. Commit messages contain rich information about the change history and the motivation of the change itself. Thus the information could bring benefits to bridge the vocabulary gap between professional developers and ordinary users.  
Another threat  arises from the selection bias of the dataset. In our experiments, we strive to reuse the same apps in the baseline work as many 
as possible. 
To reduce the noise from the raw data and bias in the result, we take the standard measures to pre-process the raw texts, and include five professional developers with over three years of mobile app development experience to solve subjective conflicts.

\medskip
\noindent{\bf External validity.}
In our case study, we deliberately selected 10 apps across different categories instead of being limited within a narrow domain. To give a fair comparison, we use a combination of multiple evaluation metrics, including both objective and subjective ones.
Similar to other empirical studies, no evidence could theoretically prove our approach can always accurately localize change files in all scenarios. But we believe that, since our approach is open to different scenarios, domain knowledge could be leveraged via new constraints and heuristics incorporated into our approach which could  improve the clustering and localization performance as well in the new dataset.

Finally, even if the underlying assumption of our work is that commit messages contribute to increase the number of common terms matching the use review vocabulary, there could be a mismatch between commit message vocabulary and user review vocabulary too. For instance, commit messages terms like bug ids (\textit{'this commit fixe the bug X'}) are not present in user reviews. Hence, for future work, we plan to investigate the potential mismatch between such vocabularies, with the goal to improve the results of our approach.

\section{Related Work}\label{related}

The concept of app store mining was introduced by Harman \etal \cite{Harman:MSR:2012} in 2012, and several researchers focused on mining mobile apps and app store data to support developers during the maintenance and evolution of mobile applications, with the goal to achieve a higher app success \cite{Martin:tse2017}.

\subsection{The Role of User Feedback Analysis in the Mobile App Success}

\textbf{App Rating  \& App Success}. Previous research widely investigated the relationship between the rating and a particular characteristic (or feature) of mobile applications \cite{Corral:2015:BCB:2825041.2825045, BavotaSE15, VasquezBBPOP13, Taba2014, TianNLH15}.  Recent research efforts have been devoted to investigating the reliability of app rating when used as a proxy to represent user satisfaction. For example, Luiz \etal~\cite{luiz2018feature} proposed a framework performing sentiment analysis on a set of relevant features extracted from user reviews. Despite the star rating was considered to be a good metric for user satisfaction, their results suggest that sentiment analysis might be more accurate in capturing the sentiment transmitted by the users. Hu \etal~\cite{hu2018studying} studied the consistency of reviews and star ratings for hybrid Android and iOS apps discovering that they are not consistent across different app markets.
Finally, Catolino \cite{catolino2018does} preliminarily investigated the extent to which source code quality can be used as a predictor of commercial success of mobile apps.

\medskip
\noindent \textbf{User Feedback Analysis \& App Success}.
Several approaches have been proposed with the aim to classify useful user reviews for app success. \textsc{AR-Miner} \cite{Chen:2014:AMI:2568225.2568263} was the first one able to classify informative reviews. Panichella \etal adopted natural language processing and text and sentiment analysis to automatically classify user reviews \cite{panichella2015can,Panichella:2016:AAR:2950290.2983938} according to a User Review Model (URM). Gu and Kim \cite{Gu:ASE15} proposed an approach that summarizes sentiments and opinions of reviews.

Following the general idea of incorporating user feedback into typical development process, Di Sorbo \etal \cite{sorbo16fse, SorboPAVC17} and Scalabrino \etal \cite{villarroel2016release,scalabrino2017listening} proposed \textsc{SURF} and \textsc{CLAP}, two approaches aiming at recommending the most important reviews to take into account while planning a new release of a mobile application. \textsc{CLAP} improves \textsc{AR-Miner} by clustering reviews into specific categories (\eg reports of security issues) and by learning from the app history (or from similar apps) which reviews should be addressed \cite{scalabrino2017listening}. \textsc{SURF} proposed a first strategy to automatically summarize user feedback in more structured and recurrent topics \cite{SorboPAVC17,Panichella18} (e.g., GUI, app pricing, app content, bugs, etc.). Finally, Palomba \etal \cite{palomba17}, inspired by the work by Scalabrino \etal, proposed ChangeAdvisor, a tool that cluster user reviews of mobile applications. In this paper we considered as baseline ChangeAdvisor since, similarly to our approach, it is based on a clustering approach for user review feedback. In evaluating our approach, we discovered that ChangeAdvisor tends to generate rather different user review clusters with the same study setting and user review data, which highlights higher reliability of our approach compared to this state-of-the-art tool.

\subsection{Information Retrieval in SE \& the Mobile Context}
Information Retrieval techniques have been widely adopted to handle several SE problems. Specifically, strategies for recovery traceability links between textual artefacts and the source code were widely studied in the past \cite{Antoniol:tse2002,DeLucia2012}. In the same way, several approaches to locating features in the source code \cite{Dit:JSEP}, and tracing informal textual documentation, such as e-mails \cite{BacchelliLR10,DiSorboASE2015,SorboPVPCG16}, forum discussions \cite{Parnin:2012, Panichella:ICPC12,VassalloICPC14}, and bug reports \cite{Saha:ase13} to the source code have been proposed. However, as previously demonstrated by Panichella \etal \cite{Panichella:2013}, the configuration used to set the clustering algorithm is an important component of topic modeling techniques used in several traceability recovery approaches, and an optimal choice of the parameters generally results in better performance.

Duan \etal \cite{duan08}  proposed a consensus-based approach to constrained clustering requirement documents.  This is a different software engineering task than ours, but both approaches employ the semi-supervised clustering technique at a high level. In Duan \etal's work \cite{duan08}, consensus clustering is firstly performed to generate an ensemble of multiple clusters, and then a voting mechanism is applied to select the constraints. In our approach, we leverage domain knowledge to help generate the constraints.
In the context of mobile computing research, two pieces of work are closer to ours. Ciurumelea \etal \cite{ciurumelea2018automated,adelina17} employed machine learning techniques for the automatic categorization of user reviews on a two-level taxonomy adopting a modified Version of Vector Space Model (VSM) to automatically link user reviews to code artefacts. Similarly, Palomba \etal \cite{palomba17} 
cluster user reviews of mobile applications and suggest the source-code artefacts to maintain. We mainly compare our approach against ChangeAdvisor \cite{palomba17} as, similar to our approach, it leverages clustering approaches for user review feedback analysis and IR-based methods for suggesting the source-code artefacts to maintain according to user change-requests. However, different from the work \cite{palomba17} and the work \cite{ciurumelea2018automated,adelina17}, the similarity score between
user reviews and source code files in our approach is calculated by a linear combination of the similarity score between the reviews and the source code 
and the  similarity score between reviews and the commit messages. 
Indeed the work \cite{palomba17,ciurumelea2018automated,adelina17} mainly relies on textual analysis techniques such as VSM and Dice to compute directly the similarity among reviews and source code files. Moreover, the word-review matrix is build on a subset of textual features, selected using PCA. This allows to select more meaningful features from user review textual content.


\section{Conclusions and Future Work}\label{conclusion}

User reviews convey client-side requirements for mobile app products. Accurate recovery of the user concerns and automatic localization of relevant source code based on these feedbacks is of great importance to facilitate rapid development. In this paper, we present an approach to localize potential change files based on user reviews for mobile applications. We conducted experiments on 10 popular mobile apps and used a comprehensive set of metrics to assess the performance of our approach. Experimental results show that our approach greatly outperforms the state-of-the-art baseline work.

In the immediate future work, we plan to develop a comprehensive environmental support for change file localization so as to give a better applicability of our approach. Moreover, our current case studies are all about open-source apps, while our future plan includes collaboration with commercial app developers and applying our approach to these industry cases.  


\section*{Acknowledgements}
This work was partially supported by the National Key R\&D Program of China (No. 2018YFB1003902), the National Natural Science Fundation of China (NSFC, No.\ 61972197), the Collaborative Innovation Center of Novel Software Technology and Industrialization, and the Qing Lan Project. T. Chen is partially supported by Birkbeck BEI School Project (ARTEFACT), NSFC grant (No.\ 61872340), and Guangdong Science and Technology Department grant (No. 2018B010107004), and an oversea grant from the State Key Laboratory of Novel Software Technology, Nanjing University (KFKT2018A16).   
\bibliographystyle{IEEEtran}
\bibliography{draft}


\end{document}